\documentclass[12pt]{article}
\usepackage{axodraw,amsmath,amssymb,amsfonts,color,graphicx,cite,color}
\usepackage{colordvi,psfrag}
\input paperdef

\graphicspath{{figs/}}

\evensidemargin \oddsidemargin
\allowdisplaybreaks

\begin{document}
\thispagestyle{empty}

\def\thefootnote{\fnsymbol{footnote}}

\begin{flushright}
DCPT/07/34 \\
IPPP/07/17 \\ 
MPP--2007--52\\
PSI--PR--07--02\\
arXiv:0705.0746 [hep-ph]
\end{flushright}

\vspace{1cm}

\begin{center}

{\Large\sc {\bf The Higgs sector of the complex MSSM\\[0.4cm] 
  at two-loop order: QCD contributions}}

\vspace{1cm}

{\sc
S.~Heinemeyer$^{1}$%
\footnote{email: Sven.Heinemeyer@cern.ch}%
, W.~Hollik$^{2}$%
\footnote{email: hollik@mppmu.mpg.de}%
, H.~Rzehak$^{3}$%
\footnote{email: Heidi.Rzehak@psi.ch}%
~and G.~Weiglein$^{4}$%
\footnote{email: Georg.Weiglein@durham.ac.uk}
}

\vspace*{.7cm}

{\sl
$^1$Instituto de Fisica de Cantabria (CSIC-UC), Santander,  Spain

\vspace*{0.1cm}

$^2$Max-Planck-Institut f\"ur Physik (Werner-Heisenberg-Institut),\\
F\"ohringer Ring 6, D--80805 M\"unchen, Germany

\vspace*{0.1cm}

$^3$Paul Scherrer Institut, W\"urenlingen und Villigen, CH--5232
Villigen PSI, Switzerland

\vspace*{0.1cm}

$^4$IPPP, University of Durham, Durham DH1~3LE, UK
}

\end{center}

\vspace*{0.1cm}

\begin{abstract}
\noindent
Results are presented for the leading two-loop contributions of
\order{\alt\als} to the masses and mixing effects in the Higgs sector
of the MSSM with complex parameters. They are obtained in the
Feynman-diagrammatic approach using on-shell renormalization. 
The full dependence on all complex phases is taken into account. 
The renormalization of the appropriate contributions of the
Higgs-boson sector and the scalar top and 
bottom sector is discussed.  
Our numerical analysis for the lightest MSSM Higgs-boson mass is based on 
the new two-loop corrections, supplemented by the full one-loop
result. 
The corrections induced by the phase variation in the scalar top sector are
enhanced by the two-loop contributions.
We find that the corresponding shift in $M_{h_1}$ can amount to~$5 \gev$.
\end{abstract}

\def\thefootnote{\arabic{footnote}}
\setcounter{page}{0}
\setcounter{footnote}{0}

\newpage


\section{Introduction}

The Higgs sector of the Minimal Supersymmetric
Standard Model (MSSM) with two scalar doublets
accommodates five physical Higgs bosons. In
lowest order these are the light and heavy $\cp$-even $h$
and $H$, the $\cp$-odd $A$, and the charged Higgs bosons $H^\pm$.
Higher-order contributions yield large corrections to the masses
and couplings, and  also induce $\cp$-violation leading to
mixing between $h,H$ and $A$ in the case of general complex SUSY breaking
parameters.

For the MSSM with real parameters (rMSSM) the status of higher-order
corrections to the masses and mixing angles in the Higgs sector is quite 
advanced~\cite{mhiggsletter,mhiggslong,mhiggsFDalbals,
bse,mhiggsEP,mhiggsRG1,
mhiggsAEC,PomssmRep,mhiggsWN,mhiggsEP5}.
In the case of the MSSM with complex parameters (cMSSM), 
the first more general investigations~\cite{mhiggsCPXgen} 
were followed by evaluations in the
effective potential approach~\cite{mhiggsCPXEP}
and with the renormalization-group-improved \onel\ effective potential
method~\cite{mhiggsCPXRG1,mhiggsCPXRG2}. 
These results have been
restricted to the corrections arising from the (s)fermion sector and
some leading logarithmic corrections from the gaugino sector%
\footnote{
The two-loop results of \cite{mhiggsEP5} can in principle also be
taken over to the cMSSM. However, no explicit evaluation or computer
code based on these results exists.
}%
. Within the Feynman diagrammatic (FD) approach the \onel\ leading $\mt^4$
corrections 
have been evaluated in \citere{mhiggsCPXFD1}. Most recently a full
one-loop calculation in the FD approach was presented~\cite{mhcMSSMlong}
(further discussions on the effect of complex phases on Higgs boson masses 
can be found in \citere{dissHR})
and implemented in 
the program \fh~\cite{mhiggslong,mhiggsAEC,feynhiggs,mhcMSSMlong},
which is publicly available.
Another public code,
{\tt CPsuperH}~\cite{cpsh}, is based on the
renormalization-group-improved effective potential
approach~\cite{mhiggsCPXRG1,mhiggsCPXRG2}. 

In this letter we improve our diagrammatic 
one-loop calculation~\cite{mhcMSSMlong}
by providing the leading \order{\alt\als} corrections
of the Higgs-boson masses and mixings in the cMSSM obtained in the FD
approach. 
Technically we calculate and renormalize the Higgs-boson self energies
taking into account the general complex parameters
of the appropriate part of the colored sector of the cMSSM.
We provide numerical examples for the lightest cMSSM Higgs-boson
mass and discuss the dependence on the phases
in the scalar top sector and on the gluino
mass parameter.
The results presented in this paper will be included in the code
\fh~\cite{mhcMSSM2Llong}.


\section{The Higgs-boson sector of the cMSSM}
\label{sec:higgs}

With the two Higgs doublets of the cMSSM decomposed in the following way,
\begin{align}
\label{eq:higgsdoublets}
\cHe = \begin{pmatrix} H_{11} \\ H_{12} \end{pmatrix} &=
\begin{pmatrix} v_1 + \tfrac{1}{\sqrt{2}} (\phi_1-i \chi_1) \\
  -\phi^-_1 \end{pmatrix}, \notag \\ 
\cHz = \begin{pmatrix} H_{21} \\ H_{22} \end{pmatrix} &= e^{i \xi}
\begin{pmatrix} \phi^+_2 \\ v_2 + \tfrac{1}{\sqrt{2}} (\phi_2+i
  \chi_2) \end{pmatrix},
\end{align}
the Higgs potential $\VHiggs$ can be
arranged as an expansion in powers of the field components, 
\begin{align}
\label{eq:higgspotential}
\VHiggs &= - T_{\phi_1} \phi_1 - T_{\phi_2} \phi_2 -
        T_{\chi_1} \chi_1 - T_{\chi_2} \chi_2 \notag \\ 
        &+ \tfrac{1}{2} \begin{pmatrix} \phi_1,\phi_2,\chi_1,\chi_2
        \end{pmatrix} 
\matr{M}_{\phi\phi\chi\chi}
\begin{pmatrix} \phi_1 \\ \phi_2 \\ \chi_1 \\ \chi_2 \end{pmatrix} +
\begin{pmatrix} \phi^-_1,\phi^-_2  \end{pmatrix}
\matr{M}_{\phi^\pm\phi^\pm}
\begin{pmatrix} \phi^+_1 \\ \phi^+_2  \end{pmatrix} + \cdots,
\end{align}
where the ellipses stand for higher powers in the Higgs-boson fields.
In \refeq{eq:higgspotential} the tadpoles appear as the coefficients
of the linear terms, and 
the bilinear terms contain the neutral and charged  mass matrices
$\matr{M}_{\phi\phi\chi\chi}$ and $\matr{M}_{\phi^\pm\phi^\pm}$. 
Tadpoles and mass matrices are conveniently rewritten in terms of 
the physical components $h,H,A,H^\pm$ and the Goldstone components
$G,G^\pm$. 
Details about the tadpole coefficients and the mass matrices can be
found in \citere{mhcMSSMlong}.  

Eq.~(\ref{eq:higgsdoublets}) introduces a 
possible new phase $\xi$ between the two Higgs doublets. 
The potential $\VHiggs$ contains the real soft breaking terms
$\tilde m_1^2$ and $\tilde m_2^2$ (with $m_1^2 \equiv \tilde m_1^2 +|\mu|^2$,
$m_2^2 \equiv \tilde m_2^2 + |\mu|^2$) and 
the generally complex soft breaking parameter $m_{12}^2$, entering the
mass matrices and tadpoles in \refeq{eq:higgspotential}.
With the help of a Peccei-Quinn
transformation~\cite{Peccei}, $\mu$ and $m_{12}^2$ can be 
redefined~\cite{MSSMcomplphasen} such that the complex phase of 
$m_{12}^2$ vanishes.
In the following we will therefore treat $m_{12}^2$ as a real
parameter, i.e.\
$|m_{12}^2| = m_{12}^2$.
Together with the requirement that 
the minimum of $\VHiggs$ is located at $v_1$ and $v_2$,
all tadpoles are zero at lowest order.

\medskip
Investigating the Higgs potential 
beyond the tree level, renormalization has to be applied to the mass matrices 
and the tadpoles, introducing counterterms according to the loop expansion
up to second order,
\begin{align}
\label{eq:deMhHAG}
\matr{M}_{hHAG} &\to \matr{M}_{hHAG} + \de \matr{M}_{hHAG}^{(1)}
                                   + \de \matr{M}_{hHAG}^{(2)} \\
\label{eq:deMHG}
\matr{M}_{H^\pm G^\pm} &\to \matr{M}_{H^\pm G^\pm} 
                          + \de \matr{M}_{H^\pm G^\pm}^{(1)} 
                          + \de \matr{M}_{H^\pm G^\pm}^{(2)} \\ 
\label{eq:deT}
T_i &\to T_i + \de T_i^{(1)} + \de T_i^{(2)}~, 
                                         \quad i = h, H, A~,
\end{align}
where the mass matrices and tadpoles are obtained from those in
\refeq{eq:higgspotential} by a rotation to the physical states.
The leading 
\order{\alt\als} contributions to the Higgs-boson self-energies
are obtained in the limit of vanishing gauge couplings 
and neglecting the dependence on the external momentum. The
bottom Yukawa coupling, appearing in the charged Higgs-boson
self-energy, is also neglected.
Since for the leading two-loop terms the Goldstone-boson parts
of \refeqs{eq:deMhHAG} and \eqref{eq:deMHG} do not contribute 
(see also the discussion in \citere{mhcMSSMlong}), 
mass renormalization at the two-loop level reduces to the counterterms 
\BE
\label{eq:deMhHA}
\de \MHp^{2(2)} \qquad {\rm and} \qquad
\de \matr{M}_{hHA}^{(2)} = 
  \begin{pmatrix}
    \de m_{h}^{2(2)}  & \de m_{hH}^{2(2)} & \de m_{hA}^{2(2)} \\[.5em]
    \de m_{hH}^{2(2)} & \de m_{H}^{2(2)}  & \de m_{HA}^{2(2)}\\[.5em]
    \de m_{hA}^{2(2)} & \de m_{HA}^{2(2)} & \de m_{A}^{2(2)}  
  \end{pmatrix}~.
\end{equation}

The renormalized Higgs-boson self-energies, denoted as
$\Hat{\Sigma}_{ij}(p^2)$ with $i,j = h,H,A,H^\pm$, are expanded into
a one-loop
and a two-loop part,  
\begin{align}
\label{eq:hSi}
\hSi_{ij}(p^2) = \hSi^{(1)}_{ij}(p^2)+
  \Hat{\Sigma}^{(2)}_{ij}(0)~.
\end{align}
The complete one-loop part has been obtained in \citere{mhcMSSMlong},
and the two-loop part is evaluated at
vanishing external momentum, as explained above.
The renormalized  two-loop self-energies
\begin{align}
\label{eq:hSi2L}
\hSi_{ij}^{(2)}(0) &= \Si_{ij}^{(2)}(0) - 
                           \de m_{ij}^{2(2)}, \quad i,j = h, H, A \\
\hSi_{H^+H^-}^{(2)}(0) &= \Si_{H^+H^-}^{(2)}(0) -
                           \de \MHp^{2(2)}
\end{align}
involve the unrenormalized Higgs-boson self-energies
$\Si_{ij}^{(2)}(0)$, containing the one-loop subrenormalization,  
and the counterterms of \refeq{eq:deMhHA}. 

\medskip
The entries $\de m_{ij}^{2(2)}$ ($i,j = h, H, A$) 
of the counterterm matrix in \refeq{eq:deMhHA}
are not all independent, but can be expressed in terms of 
$\de\MHp^{2(2)}$ and $\de T_i^{(2)}$. As explained e.g.\ 
in \citere{mhiggslong},
$\MHp^2$ and $T_i$ are the only independent parameters in the 
Higgs potential that have to be renormalized for the evaluation of the
\order{\alt\als} terms. Correspondingly, it is sufficient to
impose renormalization conditions for the tadpoles and for
the charged Higgs-boson mass:
\begin{itemize}
\item The tadpoles are fixed by the requirement that the
  minimum of the Higgs potential is not shifted, yielding at the
  two-loop level
\BEA
T_i^{(2)} + \de T_i^{(2)} = 0 &\Rightarrow&
  \de T_i^{(2)} = - T_i^{(2)}~, ~i = h, H, A~.
\EEA
\item The mass square of the charged Higgs boson, $\MHp^2$,
is fixed by an on-shell condition yielding the
following counterterm at the two-loop level:
\BEA
\label{eq:deM}
\re\hSi_{H^+H^-}(0) = 0 &\Rightarrow& 
  \de\MHp^{2(2)} = \Si_{H^+H^-}^{(2)}(0)~.
\EEA
\end{itemize}
With these, the counterterms for the neutral mass matrix 
are now determined in the following way,
\begin{align}
\de m_{h}^{2(2)} &= c_{\al-\be}^2\, \de \MHp^{2(2)} 
+ \frac{e\, s_{\al - \be}}{4 \MZ \cw \sw} 
  \Bigl[ \bigl(s_{\al - 2 \be} - 3 s_\al \bigr) \de T_{\phi_1}^{(2)} 
   + \bigl(c_{\al - 2 \be} + 3 c_\al \bigr)\de T_{\phi_2}^{(2)}\Bigr]~,\\ 
\de m_{hH}^{2(2)} &= \frac{s_{2\al-2\be}}{2}\de\MHp^{2(2)}
+ \frac{e\, c_{\al - \be}}{4 \MZ \cw \sw} 
  \Bigl[ \bigl(s_{2 \al - \be} + c_{2\al - 2\be} s_\be \bigr) 
         \de T_{\phi_1}^{(2)} 
   - \frac{c_{2 \al - 3 \be} + 3 c_{2 \al -\be}}{2}
         \de T_{\phi_2}^{(2)}\Bigr]~, \\
\de m_{H}^{2(2)} &= s^2_{\al -\be}\, \de\MHp^{2(2)} 
  +\frac{e\, c_{\al - \be}}{4\MZ \cw \sw} 
   \Bigl[ \bigl(c_{\al - 2 \be} - 3 c_\al \bigr) \de T_{\phi_1}^{(2)} 
        - \bigl(s_{\al - 2 \be} + 3 s_\al \bigr) \de T_{\phi_2}^{(2)}\Bigr]~,\\
\de m_{AH}^{2(2)} &= - \frac{e\, c_{\al - \be}}{2 \MZ \cw \sw} 
  \de T_A^{(2)}~,\\
\de m_{Ah}^{2(2)} &= \frac{e\, s_{\al - \be}}{2 \MZ \cw \sw} \de T_A^{(2)}~,\\
\de m_{A}^{2(2)} &= \de\MHp^{2(2)}~. 
\label{deVAA}
\end{align}
We have used $s_x \equiv \sin(x)$, $c_x \equiv \cos(x)$ as
abbreviations. 
The angle $\al$ diagonalizes the $\phi_1\phi_2$ mass matrix at
tree-level, $T_{\phi_1}$ and $T_{\phi_2}$ denote the $H$ and $h$
tadpoles, respectively, in the limit of $\al \to 0$, which are the
tadpoles in \refeq{eq:higgspotential}.

\begin{figure}[htb!]
\begin{center}
\includegraphics[width=0.9\linewidth]{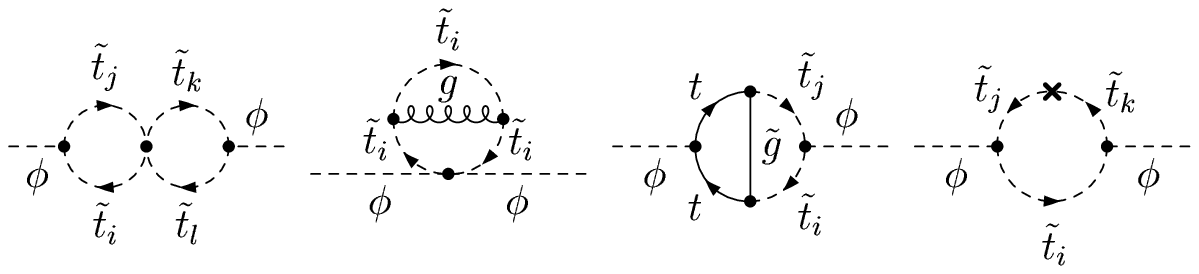}
\caption{Examples of generic \twol\ diagrams and diagrams with counterterm
insertion for the Higgs-boson self-energies
($\phi = h, H, A$;
$\;i,j,k,l = 1,2$).}
\label{fig:fd_hHA}
\end{center}
\end{figure}
\begin{figure}[htb!]
\begin{center}
\includegraphics[width=0.9\linewidth]{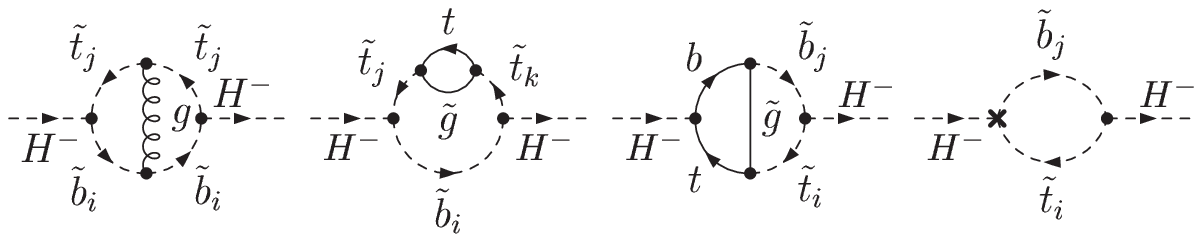}
\caption{Examples of generic \twol\ diagrams and diagrams with
  counterterm insertion for the charged Higgs-boson self-energy
($i,j,k = 1,2$).}
\label{fig:fd_Hpm}
\end{center}
\end{figure}
\begin{figure}[htb!]
\begin{center}
\includegraphics[width=0.9\linewidth]{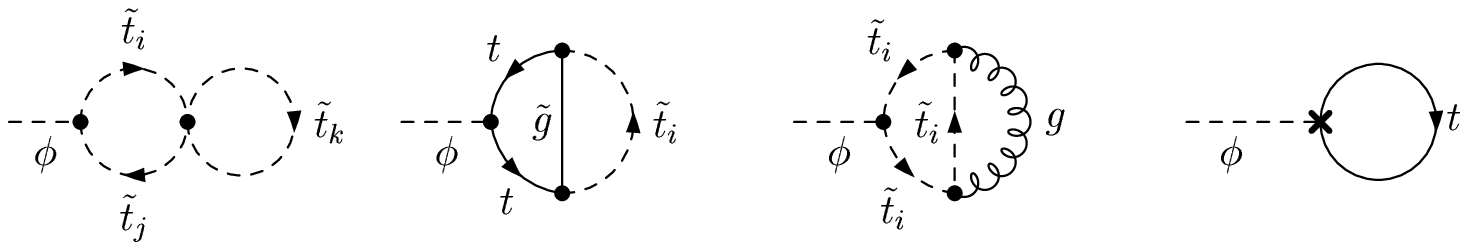}
\caption{Examples of generic \twol\ diagrams and diagrams with
  counterterm insertion for the Higgs-boson tadpoles, 
($\phi = h, H, A$;
$\;i,j,k = 1,2$).}
\label{fig:fd_TP}
\end{center}
\end{figure}

\smallskip
The calculation of the unrenormalized self-energies and tadpoles
at \order{\alt\als} requires the evaluation of genuine two-loop diagrams
and one-loop graphs with counterterm insertions.
Example diagrams for the neutral Higgs-boson self-energies
are depicted in \reffi{fig:fd_hHA}, and for 
the charged Higgs boson in \reffi{fig:fd_Hpm}. 
Examples for
the tadpole diagrams are displayed in \reffi{fig:fd_TP}. 
The complete set of contributing Feynman diagrams 
has been generated with the
program {\tt FeynArts}~\cite{feynarts};
tensor reduction and the evaluation of traces was done with 
support of the programs {\tt OneCalc} and {\tt TwoCalc}~\cite{twocalc},
yielding analytic expressions in terms of 
the scalar one-loop functions $A_0, B_0$~\cite{oneloop} 
and two-loop vacuum integrals~\cite{twoloop}.
The numerical evaluation was performed with the help of the program 
{\tt LoopTools}~\cite{looptools}.

\smallskip
The renormalized self-energies determine the dressed propagators of the
Higgs fields, from which masses and mixing properties at higher order are
derived. The self-energies have an impact on the location of the 
poles and thus on the Higgs particle masses, which are in general
different from their tree-level values. Only the charged 
Higgs boson mass $\MHp$ is not shifted, owing to the on-shell
renormalization condition~(\ref{eq:deM}).

The non-diagonal self-energies are responsible for 
mixing in the neutral Higgs system.
In the presence of complex parameters all
three neutral $\cp$ eigenstates $h,H,A$ can mix. 
The $3 \times 3$ propagator matrix, $\De_{hHA}(p^2)$, 
is obtained by inverting the renormalized irreducible two-point function,
\begin{equation}
\De_{hHA}(p^2) = - \left(\hat{\Gamma}_{hHA}(p^2)\right)^{-1} ,
\label{eq:propagator}
\end{equation}
where
\begin{align}
\label{eq:invprophiggs}
  \hat{\Gamma}_{hHA}(p^2) &= i \left[p^2 \unity - \matr{M}_{\mathrm{n}}(p^2)
                               \right], \\[.5em]
  \matr{M}_{\mathrm{n}}(p^2) &=
  \begin{pmatrix}
    \mh^2 - \ser{hh}(p^2) & - \ser{hH}(p^2) & - \ser{hA}(p^2) \\
    - \ser{hH}(p^2) & \mH^2 - \ser{HH}(p^2) & - \ser{HA}(p^2) \\
    - \ser{hA}(p^2) & - \ser{HA}(p^2) & \mA^2 - \ser{AA}(p^2)
  \end{pmatrix}. 
\label{eq:Mn}
\end{align}
The masses of the three Higgs-boson mass eigenstates, $h_1$, $h_2$, $h_3$,
ordered according to $M_{h_1} \leq M_{h_2} \leq M_{h_3}$,
are given by the real parts of the poles of 
$\De_{hHA}(p^2)$ or, equivalently, of the roots 
of the determinant of the two-point vertex function,
${\rm det}[\hat{\Gamma}_{hHA}(p^2)]=0$.
The quantities $\mh,\mH,\mA$ in \refeq{eq:Mn} are the 
masses of $h,H,A$ at the tree-level, respectively.


\section{The colored sector of the cMSSM}
\label{sec:color}

For the evaluation of the \order{\alt\als} two-loop contributions
to the tadpoles and self-energies, a renormalization of the one-loop
contributions from the 
scalar top ($\Stop$) and bottom ($\Sbot$) sector 
is needed, giving rise to the counterterms for one-loop subrenormalization
(see \reffis{fig:fd_hHA}--\ref{fig:fd_TP}). 
The bilinear part of the $\Stop$ and $\Sbot$ Lagrangian,
\begin{align}
\cL_{\Stop/\Sbot\text{ mass}} &= - \begin{pmatrix}
{{\tilde{t}}_{L}}^{\dagger}, {{\tilde{t}}_{R}}^{\dagger} \end{pmatrix}
\matr{M}_{\tilde{t}}\begin{pmatrix}{\tilde{t}}_{L}\\{\tilde{t}}_{R}
\end{pmatrix} - \begin{pmatrix} {{\tilde{b}}_{L}}^{\dagger},
{{\tilde{b}}_{R}}^{\dagger} \end{pmatrix}
\matr{M}_{\tilde{b}}\begin{pmatrix}{\tilde{b}}_{L}\\{\tilde{b}}_{R} 
\end{pmatrix}~,
\end{align}
contains the stop and sbottom mass matrices
$\matr{M}_{\tilde{t}}$ and $\matr{M}_{\tilde{b}}$,
given by 
\begin{align}\label{Sfermionmassenmatrix}
\matr{M}_{\tilde{q}} &= \begin{pmatrix}  
 M_L^2 + m_q^2 + M_Z^2 c_{2 \beta} (T_q^3 - Q_q \sw^2) & 
 m_q \Xq^* \\[.2em]
 m_q \Xq &
 M_{\tilde{q}_R}^2 + m_q^2 +M_Z^2 c_{2 \beta} Q_q \sw^2
\end{pmatrix}~, \quad q = t,\, b~, \\
{\rm with} &\mbox{} \non \\
\Xq &= \Aq - \mu^*\kappa~, \qquad \kappa = \{\cot\beta, \tan\beta\} 
        \quad {\rm for} \quad q = t,b~.
\end{align}
$Q_{{q}}$ and $T_q^3$ denote charge and isospin of $q$, and
$A_q$ is the trilinear soft-breaking parameter.
The mass matrix can be diagonalized with the help of a unitary
 transformation ${\matr{U}}_{\tilde{q}}$, which can be parametrized
 by a mixing angle $ {\theta}_{\tilde{q}}$ and a phase
 $\varphi_{\tilde q}$,
\begin{align}\label{transformationkompl}
\matr{D}_{\tilde{q}} &= 
\matr{U}_{\tilde{q}}\, \matr{M}_{\tilde{q}} \, 
{\matr{U}}_{\tilde{q}}^\dagger = 
\begin{pmatrix} \msqe^2 & 0 \\ 0 & \msqz^2 \end{pmatrix}\,,\quad\
{\matr{U}}_{\tilde{q}}= 
\begin{pmatrix} U_{\tilde{q}_{11}}  & U_{\tilde{q}_{12}} \\  
                U_{\tilde{q}_{21}} & U_{\tilde{q}_{22}}  \end{pmatrix}
= \begin{pmatrix} 
  \cos {\theta}_{\tilde{q}} & 
  e^{i {\varphi}_{\tilde{q}}} \sin {\theta}_{\tilde{q}} \\ 
  - e^{- i{\varphi}_{\tilde{q}}} \sin {\theta}_{\tilde{q}} & 
  \cos {\theta}_{\tilde{q}} \end{pmatrix}~. 
\end{align}
The mass eigenvalues depend only on $|\Xq|$.

\medskip
Taking into account complex phases, the renormalization in the
$\Stop$~sector is somewhat more involved than in the case of real
parameters~\cite{mhiggslong,mhiggsFDalbals,hr}.
In the cMSSM the $\Stop$~sector is described in terms of five real
parameters (where we assume that $\mu$ and $\tb$ are defined via other
sectors): the real soft SUSY-breaking
parameters $M_L^2$ and $M_{{\tilde{t}}_R}^2$, the absolute value and
complex phase of the trilinear coupling, $A_t = |A_t| e^{i \varphi_{A_t}}$, 
and the top Yukawa coupling $\lambda_t$ that can be chosen to be real. 
Instead of the quantities
$M_L^2$, $M_{{\tilde{t}}_R}^2$ and $\lambda_t$, in the on-shell scheme
applied in this paper we choose the
on-shell squark masses $\mste^2$, $\mstz^2$ and the top-quark mass
$\mt$ as independent parameters.

\smallskip
The following renormalization conditions are imposed:
\begin{itemize}
\item[(i)] The top-quark mass is defined on-shell, yielding the one-loop
  counterterm $\de \mt$:
\begin{align}\label{dmt}
\de  \mt = \frac{1}{2} \mt \bigl(
 \widetilde{\text{Re}}{\Si}_t^L (\mt^2) 
+ \widetilde{\text{Re}} {\Si}_t^R (\mt^2) 
+ 2 \widetilde{\text{Re}}{\Si}_t^S (\mt^2)\bigr) ~,
\end{align}
referring to the Lorentz decomposition of the self energy ${\Si}_{t}$
\begin{align}
{\Si}_{t} (k) &= \not\! k {\omega}_{-}
{\Si}_t^L (k^2) +\not\! k {\omega}_{+}
{\Si}_t^R (k^2) + \mt {\Si}_t^S (k^2) + \mt \gamma_5
{\Si}_t^{PS} (k^2) \label{decomposition}
\end{align}
into a left-handed, a right-handed, a
scalar and a pseudoscalar part, ${\Si}_t^L$, ${\Si}_t^R$, ${\Si}_t^S$
and ${\Si}_t^{PS}$, respectively.  
$\widetilde{\text{Re}}$ denotes the real part with respect to
contributions from the loop integral, but leaves the complex
couplings unaffected.
\item[(ii)]
The stop masses are also determined via on-shell
conditions~\cite{mhiggslong,hr}, yielding  
\begin{align}\label{dmst}
\de  m_{\tilde{t}_i}^2 &= 
\widetilde{\text{Re}}\Si_{\tilde{t}_{ii}}(m_{{\tilde{t}}_{i}}^2)
\quad\ \text{with} \quad\ i = 1,\,2~.
\end{align}
\item[(iii)]
The third condition affects the stop mixing angle and phase, or
equivalently, the $A_t$ parameter.
Rewriting the squark mass matrix in terms of the mass eigenvalues and
the mixing angle and phase
using \refeq{transformationkompl},
\begin{align}
\label{CTexpansion}
{\matr{M}}_{\Stop} &= \begin{pmatrix} 
  \cos^2\tst \mste^2 + \sin^2\tst \mstz^2 & 
  e^{i {\varphi}_{\Stop}} \sin\tst \cos\tst(\mste^2 - \mstz^2) \\ 
  e^{- i {\varphi}_{\Stop}} \sin\tst \cos \tst(\mste^2 - \mstz^2) & 
  \sin^2\tst \mste^2 + \cos^2\tst \mstz^2 
\end{pmatrix} \, ,
\end{align}
yields the counterterm matrix $\delta {\matr{M}}_{\Stop}$ 
by introducing counterterms $\delta\mste^2, \delta\mstz^2$
for the masses and $\delta\tst, \delta\varphi_{\Stop}$
for the angles.
For the case of the real MSSM one obtains 
the counterterm for the mixing angle,
\begin{align}
\label{Ydef}
\mbox{rMSSM:} \qquad
 (m^2_{\Stop_1} - m^2_{\Stop_2}) \, \de \theta_{\Stop} =
 [\matr{U}_{\Stop}\, \de\matr{M}_{\Stop}\, \matr{U}_{\Stop}^\dagger]_{12}
 \equiv \de Y_{\Stop} \, ,
\end{align}
for which the following renormalization condition has been 
used~\cite{mhiggsFDalbals,hr}:
\begin{align}
\label{Yrenormalization}
\mbox{rMSSM:} \qquad
\de Y_{\Stop} = \frac{1}{2}\, [  
 {\text{Re}}{\Si}_{\Stop_{12}}(\mste^2)  +
 {\text{Re}}{\Si}_{\Stop_{12}}(\mstz^2)] \, .
\end{align}
Generalizing \refeq{Ydef} to the complex case, we obtain
\begin{align}
\de Y_{\Stop} = 
 [\matr{U}_{\Stop}\, \de\matr{M}_{\Stop}\, \matr{U}_{\Stop}^\dagger]_{12} = 
 (m^2_{\Stop_1} - m^2_{\Stop_2}) \, 
  e^{i\varphi_{\Stop}} \, (\de \theta_{\Stop}  + 
  i \sin\theta_{\Stop} \cos\theta_{\Stop} \,\, \de\varphi_{\Stop} )~,
\end{align}
and impose, as a generalization of \refeq{Yrenormalization}, the condition
\begin{align}
\de Y_{\Stop} = \frac{1}{2}\, [
 \widetilde{\text{Re}}{\Si}_{\Stop_{12}}(\mste^2)  +
 \widetilde{\text{Re}}{\Si}_{\Stop_{12}}(\mstz^2)] \, ,
\end{align}
which now corresponds to two separate conditions for the real and
imaginary part, or for 
$\de\theta_{\Stop}$ and $\de\varphi_{\Stop}$, respectively.

We adopt a scheme where $|\At|$ and $\phiat$ are chosen as independent
parameters. 
The two sets of parameters $\tst$, $\varphi_{\Stop}$ and $|\At|$,
$\phiat$ are mutually related via \refeq{Sfermionmassenmatrix} 
and \refeq{CTexpansion}. The off-diagonal entries of the corresponding
counterterm matrices yield
\BE
(\At^* - \mu \cot\beta)\, \de \mt + \mt\, \de \At^* =
U^*_{\Stop_{11}} U_{\Stop_{12}}
(\de \mste^2 - \de \mstz^2) +
U^*_{\Stop_{11}} U_{\Stop_{22}} \de Y_t + U_{\Stop_{12}}
U^*_{\Stop_{21}} \de Y_t^*~.
\end{equation}
As a result, we obtain for $\de |\At|$ and  $\de \varphi_{\At}$
\begin{align}
\label{dAtbetrag1}
\de |\At| &=  \frac{1}{\mt}\text{Re}[e^{i \phiat} K_t]~,   
\\\label{Atphase1} 
\de \phiat &= - \frac{1}{\mt |\At|}
                   \text{Im}[e^{i \phiat} K_t]~,
\end{align}
with
\begin{align}
K_t &= \nonumber
- (\At^* - \mu \cot \beta)\de \mt 
+ U_{\Stop_{11}}^* U_{\Stop_{12}}
(\de \mste^2 - \de \mstz^2) 
 +  U_{\Stop_{11}}^{*} U_{\Stop_{22}} \de Y_t
+ U_{\Stop_{12}} U_{\Stop_{21}}^{*} \de Y_t^*~.
\end{align}

\end{itemize}

In the scalar bottom sector, we also encounter 
five real parameters (with $\mu$ and $\tb$ defined via other sectors):
the real soft-breaking 
parameters $M_L^2$ and $M_{{\tilde{b}}_R}^2$, the absolute value and
phase of the trilinear coupling $A_b$, and the bottom 
Yukawa coupling $\lambda_b$ that can be chosen to be real (for the set
of corrections presented in this paper $\la_b$ does not enter, as
explained above).
SU$(2)$ invariance requires the ``left-handed'' soft-breaking 
parameters in the stop and the sbottom sector to be
identical (denoted as $M_L^2$).
In the evaluation of the \order{\alt\als} contributions to
the Higgs-boson self-energies, the counterterms of the sbottom sector 
appear only in the self-energy of the charged Higgs boson. 
In our approximation, where the 
$b$-quark mass is neglected, 
$\SbotL$ and
$\SbotR$ do not mix, 
and $\SbotR$ decouples and does not contribute.
The charged Higgs-boson self-energy thus depends only on a single
parameter of the sbottom sector, which can be chosen as the squark mass
$m_{\tilde{b}_L}$. The parameter $\msbl$ should be regarded simply as
the upper left entry in the $\Sbot$~mass matrix, not as a physical
$\Sbot$~pole mass (see also \citere{mhiggsFDalbals, hr}).
By means of SU$(2)$ invariance, the corresponding mass  counterterm
is already determined:
\begin{align}
\label{dmsb}
\de \msbl^2 = 
 |U_{\Stop_{11}}|^2 \de \mste^2
 + |U_{\Stop_{12}}|^2 \de \mstz^2
 - U_{\Stop_{12}}^* U_{\Stop_{22}} \de Y_t
 - U_{\Stop_{12}} U_{\Stop_{22}}^* \de Y_t^*
- 2 \mt \de \mt~.
\end{align} 

With the set of renormalization constants determined in \refeqs{dmt},
(\ref{dmst}), (\ref{dAtbetrag1}), (\ref{Atphase1}) and (\ref{dmsb})
the counterterms for the diagonal and non-diagonal (s)quark
self-energies as well as 
for all Higgs-boson--(s)quark vertices  
are at our disposal for the one-loop subrenormalization.
An explicit list of the counterterms will be provided in a detailed
forthcoming publication~\cite{mhcMSSM2Llong}.

\medskip
Finally, at \order{\alt\als} gluinos appear as
virtual particles at the two-loop level (hence, no renormalization is
needed). The corresponding 
soft-breaking gluino mass parameter $M_3$ is in general
complex, 
\BE
M_3 = |M_3| e^{i \phigl} \quad (\text{with the gluino mass}~\mgl = |M_3|)~.
\end{equation}
The phase can be absorbed by a redefinition of the gluino Majorana
spinor such that it appears only in the gluino couplings, but not in
the mass term.


\section{Numerical results}
\label{sec:numeval}

We illustrate the effects of the two-loop contributions in terms 
of the mass of the lightest neutral Higgs boson, $\MHe$,
evaluated on the basis of \refeq{eq:Mn} with the entries from \refeq{eq:hSi}.
The results for physical observables are affected only
by certain combinations of the complex phases.
In particular, the \order{\alt\als} corrections presented in this paper 
depend only on the combinations~\cite{MSSMcomplphasen,SUSYphases}
\BE
\mu\,\At\,\left(m_{12}^2\right)^* \mbox{~~~and~~~} \At\,M_3^* ~ .
\label{eq:complphases}
\end{equation}
As discussed above, we have transformed our parameters such that the
complex phase of $m_{12}^2$ vanishes. 
Therefore our two-loop results depend on the phases of the parameters
$\At$, $\mu$ and $M_3$, which we denote as $\phiat$, $\phimu$ and $\phigl$,
respectively. We do not consider the variation of complex phases that
enter only via one-loop contributions.

In the context of a detailed phenomenological analysis of the cMSSM
parameter space the existing constraints on $\cp$-violating parameters
from experimental bounds for the electric dipole moments
(EDMs)~\cite{pdg,plehnix} are of interest. 
While SM contributions enter 
only at the three-loop level, due to its
complex phases the cMSSM can contribute to the EDMs already at one-loop order.
The complex phases appearing in the cMSSM are experimentally 
constrained by their contribution to the EDMs of
heavy quarks~\cite{EDMDoink}, of the electron and 
the neutron (see \cite{EDMrev2,EDMPilaftsis} and references therein), 
and of deuterium~\cite{EDMRitz}. 
One finds that in particular the phase $\varphi_\mu$ is tightly
constrained (in the convention where the phase of the gaugino mass
parameter $M_2$ is set to zero). 
The bounds on the phases of the third-generation trilinear couplings,
on the other hand, are much weaker.

Since the complex phases appear in our two-loop result only in the
combinations given in \refeq{eq:complphases}, we can conveniently choose
$\phimu = 0$, so that in our numerical analysis only $\phiat$ and
$\phigl$ are varied. 
In order to illustrate the possible effects of complex phases 
we will show below results for $\phigl$, $\phiat$
varied over the full parameter range. 

Our numerical analysis has been performed for the following set of
parameters (if not indicated differently):
\BEA
&& \msusy = 1000 \gev, \; |\At| = |\Ab| = |\Atau| = 1000 \gev, \;
   \phiab = \phiatau = 0, 
   \; \non \\
&& \mu = 1000 \gev, \; M_2 = 500 \gev, \; 
   M_1 = (5\sw^2)/(3\cw^2)\,M_2 , \;
\mgl = 1000 \gev, \non \\
&& \MHp = 500 \gev, \; \tb = 10, %
                                                \; \mt = 174.3 \gev~.
\label{parameters}
\EEA
$\msusy$ denotes the diagonal soft SUSY-breaking parameters in the
sfermion mass matrices that are chosen to be equal to each other.
We do not consider higher values of $\tb$, which in general enhance
the SUSY contributions to the EDMs.

\begin{figure}[!t]
\includegraphics[width=0.99\linewidth]{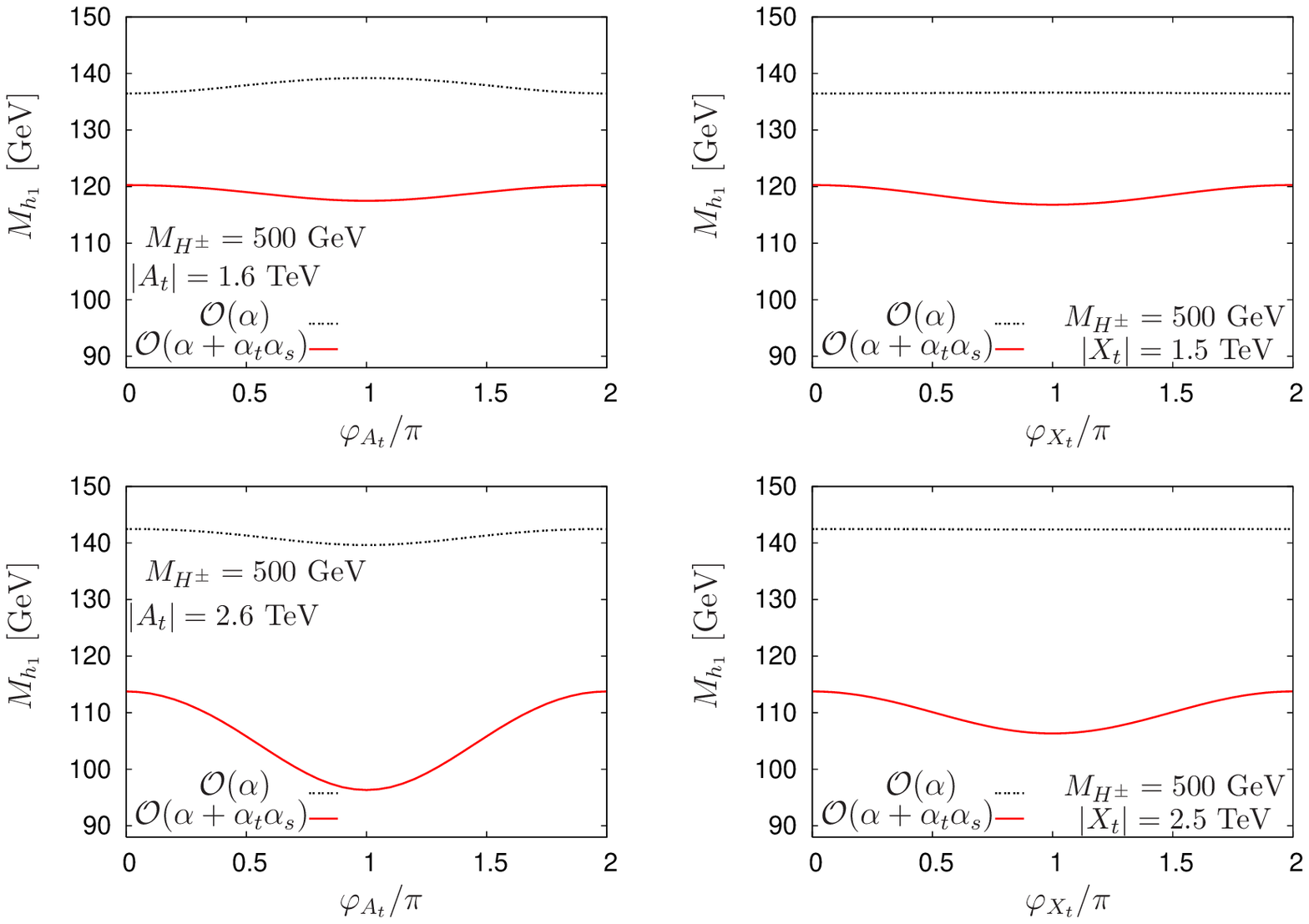}
\caption{
In the left column $\MHe$ is given as a function of $\phiat$ 
for $|\At| = 1.6 \tev$ (upper plot) and $|\At| = 2.6 \tev$ (lower plot).
In the right column $\MHe$ is given as a function of $\phixt$ 
for $|\Xt| = 1.5 \tev$ (upper plot) and $|\Xt| = 2.5 \tev$ (lower plot).
The other parameters are as given in \refeq{parameters}
and $\phigl = 0$. 
The one-loop results (dashed line) are compared with the results
including the \order{\alt\als} corrections (solid line).
}
\label{fig:Mh1phiAtphiXt}
\end{figure}

We first discuss 
the dependence of $\MHe$ on the phase in the
scalar top sector. Since the leading one-loop result in the limit 
$\MHp \gg \MZ$ depends only on the absolute value 
$|\Xt| \equiv |\At - \mu^*/\tb|$ (implying that only the
combination $\phiat + \phimu$ enters, in accordance with
\refeq{eq:complphases}), it is useful to analyze the dependence of the
result on $\phiat$ as well as on $\phixt$.%
\footnote{
It should be noted that the variation of $\MHe$ with $\phixt$ can be
substantial for small values of $\MHp$~\cite{dissHR}.
}
In \reffi{fig:Mh1phiAtphiXt} we show the lightest
Higgs-boson mass as a function of $\phiat$ (left) and of $\phixt$
(right) for $|\Xt| = 1.5 \tev$ (upper row) and $|\Xt| = 2.5 \tev$
(lower row). $|\At|$ is chosen such that for vanishing phases
it is equal in the left and right plot of each row. A variation
of $\phiat$ for fixed $\mu$ and $\tb$ changes the absolute value of
$\Xt$ and thus the masses of the scalar top quarks. Changing $\phixt$,
on the other hand, leaves the masses of the scalar tops invariant
(see \refse{sec:color}), but changes $\At$. 
Therefore, in the right plots the $\Stop$~masses are constant 
($\mste = 770 \gev$ and $\mstz = 1210 \gev$). 
We compare in \reffi{fig:Mh1phiAtphiXt} the one-loop 
result for $\MHe$ (dotted line) with the new result that includes the
\order{\alt\als} contributions (solid line). 

\begin{figure}[htb!]
\includegraphics[width=0.99\linewidth]{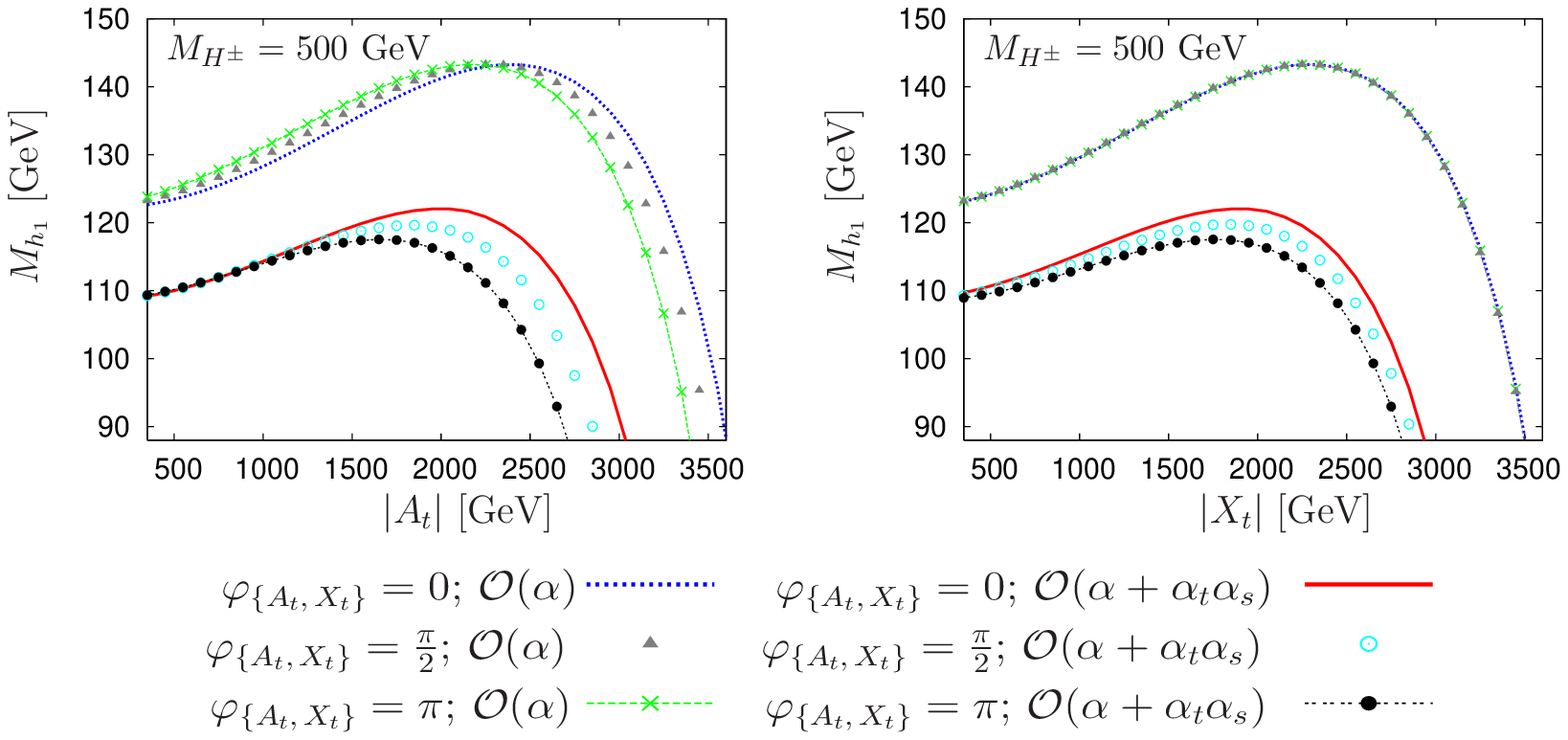}
\caption{
$\MHe$  as a function of $|\At|$ (left plot) and $|\Xt|$
(right plot) for $\phixt, \phiat = 0, \pi/2, \pi$. 
The other parameters are as given in \refeq{parameters}
and $\phigl = 0$. 
The one-loop results are compared with the results
including the \order{\alt\als} corrections.
}
\label{fig:Mh1AtXt}
\end{figure}

\begin{figure}[htb!]
\includegraphics[width=0.99\linewidth]{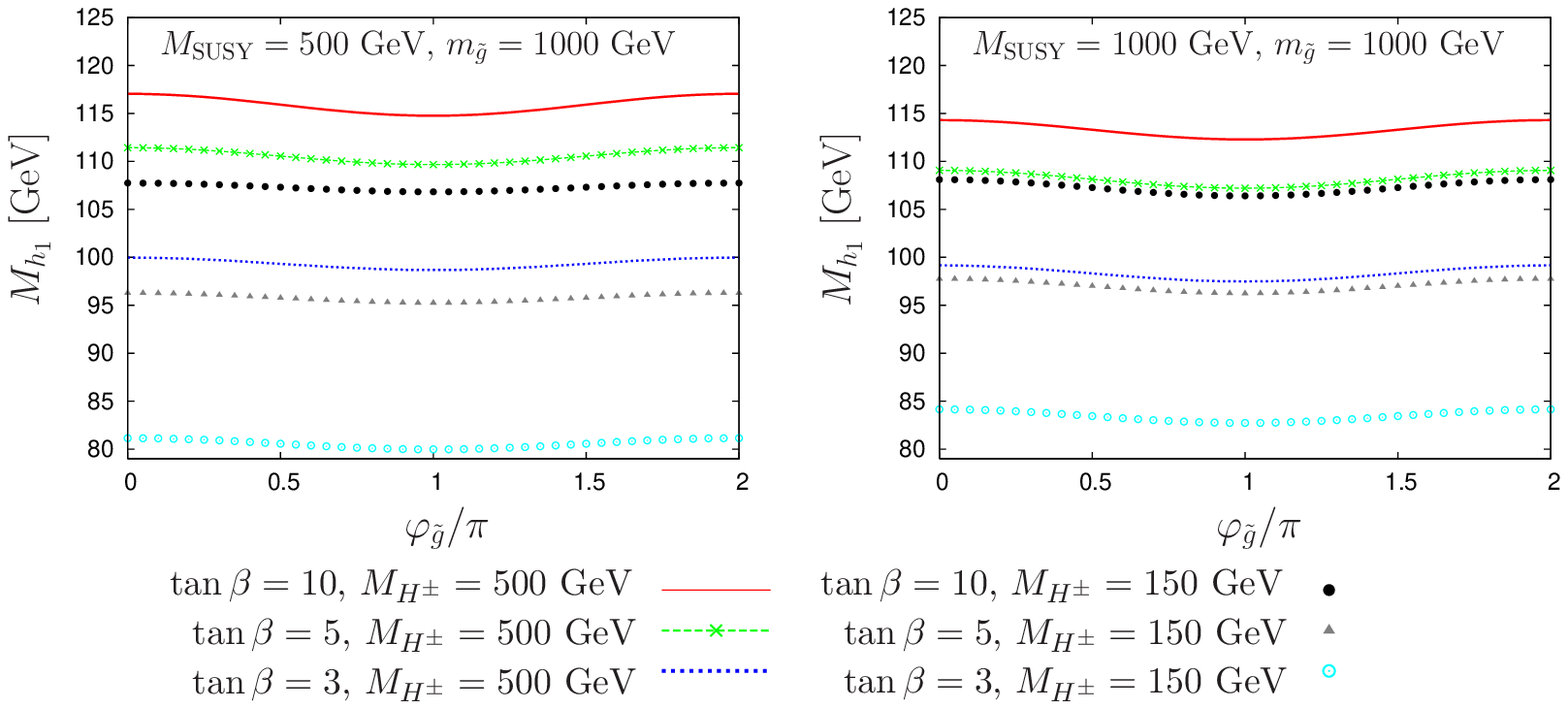}
\caption{
  $\MHe$ at \order{\alt\als} as a function of $\phigl$ for
  $\msusy = 500 \gev$ 
  (left) and $\msusy = 1000 \gev$ (right) with $\MHp = 150, 500 \gev$,
  $\tb = 3, 5, 10$ and $\phiat = 0$.
}
\label{fig:Mh1phigl}
\end{figure}

\begin{figure}[htb!]
\includegraphics[width=0.99\linewidth]{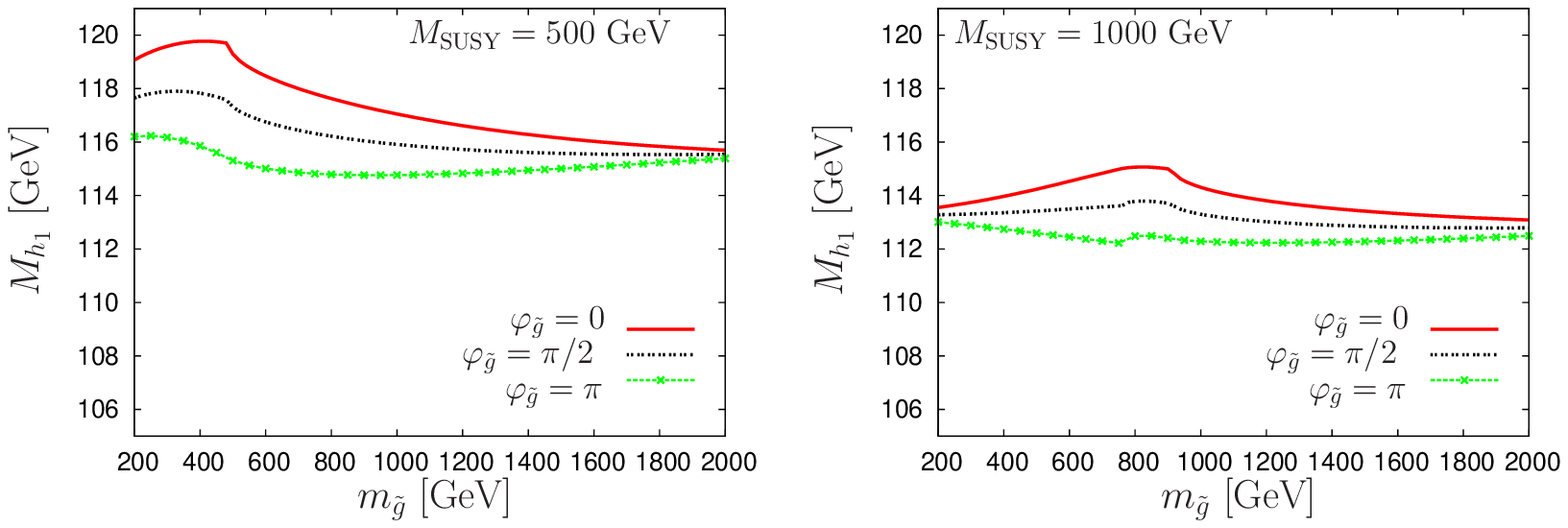}
\caption{
  $\MHe$ at \order{\alt\als} as a function of $\mgl$ for
  $\msusy = 500 \gev$ 
  (left) and $\msusy = 1000 \gev$ (right) with $\phigl = 0, \pi/2, \pi$
  and $\phiat = 0$.
}
\label{fig:Mh1mgl}
\end{figure}

\reffi{fig:Mh1phiAtphiXt} shows that the two-loop contributions lead
to a reduction of $\MHe$ of $\sim 20 \gev$, in accordance with the results 
in the rMSSM~\cite{mhiggsletter,mhiggslong}.
The dependence on the complex phases $\phiat$ and $\phixt$ is much more
pronounced in the two-loop result than in the one-loop case. This fact
can easily be understood from the discussion above: for the relatively
large $\MHp$ chosen in \reffi{fig:Mh1phiAtphiXt} the one-loop result is
dominated by contributions involving only the absolute value $|\Xt|$.
Therefore the dependence of the one-loop result on $\phixt$ is very weak
(right column of \reffi{fig:Mh1phiAtphiXt}), while the dependence on 
$\phiat$ arises to good approximation only from its effect on $|\Xt|$.
At the two-loop level, on the other hand, the contributions with
internal gluinos depend on the phase of $(\At \, M_3^*)$, see
\refeq{eq:complphases}. This induces an asymmetry of the leading corrections 
to $\MHe$ with respect to $\Xt$ (see \citeres{mhiggslong,bse} for
a discussion in the rMSSM).

The impact of the phases $\phiat$, $\phixt$ is obviously enhanced 
for larger values of $|\At|$ and $|\Xt|$.
In \reffi{fig:Mh1AtXt} we show the dependence of $\MHe$ on
$|\At|$ (left) and $|\Xt|$ (right) for $\phiat, \phixt = 0, \pi/2, \pi$,
respectively. Concerning the dependence on $|\Xt|$, 
at the one-loop level the results are indistinguishable for the three values
of $\phixt$, in agreement with the one-loop results in
\reffi{fig:Mh1phiAtphiXt}. Varying $\phiat$, on the other hand, results
in a shift of the position of the maximum of $\MHe$ in the one-loop
result (left plot). At the two-loop level, the position and size of 
the maximum value of $\MHe$ is significantly affected both by 
$\phiat$ and $\phixt$, in accordance with the discussion above.

We now investigate the dependence of $\MHe$ on the phase of the gluino
mass parameter, keeping the phase of $\At$ fixed at $\phiat = 0$. 
This means 
that only the second term in \refeq{eq:complphases} is affected by the
phase variation, while so far we had studied the combined effect of both
terms in \refeq{eq:complphases}.
\reffi{fig:Mh1phigl}  displays the variation of
$\MHe$ with $\phigl$ for $\msusy = 500 \gev$ (left) and 
$\msusy = 1000 \gev$ (right). $\MHp$ is set to $150, 500 \gev$, and 
$\tb = 3, 5, 10$. The dependence on the gluino phase (for $\phiat = 0$)
is relatively weak for the set of parameters chosen in \reffi{fig:Mh1phigl},
yielding shifts in $\MHe$  below $\sim 2 \gev$.  
For larger $\MHp$ the dependence
is slightly stronger than for small $\MHp$ values. In all cases a
minimum of $\MHe$ is reached for $\phigl = \pi$. 
Larger effects of the phase of the gluino mass parameter than the ones shown
in the example of \reffi{fig:Mh1phigl} would occur for larger values of
$|\At|$, as a consequence of \refeq{eq:complphases}. 

In \reffi{fig:Mh1mgl} the result for $\MHe$ is shown as a
function of $\mgl$ for $\phigl = 0, \pi/2, \pi$ (and $\phiat = 0$). 
$\msusy$ is set to 
$500 \gev$ (left) and $1000 \gev$ (right). The phase dependence is
strongest around the thresholds $\mgl = \mste - \mt$ and 
$\mgl = \mstz - \mt$. For the chosen set of parameters the thresholds
correspond to $\mgl = 177 \gev$ (not shown) and $\mgl = 487 \gev$ for
$\msusy = 500 \gev$, and to $\mgl = 760 \gev$ and $\mgl = 915 \gev$ for
$\msusy = 1000 \gev$. The change in $\MHe$ induced by the phase
variation can amount up to $4 \gev$ in the threshold area for the parameters
chosen in \reffi{fig:Mh1mgl}.


\section{Conclusions}
\label{sec:conclusions}

We have presented results for the leading \order{\alt\als} contributions
to the dressed Higgs-boson propagators in the MSSM with
complex parameters, obtained in the
Feynman-diagrammatic approach using an on-shell type renormalization scheme.
In the Higgs sector a two-loop renormalization has to be carried out for 
the mass of
the charged Higgs boson and the three tadpoles. The renormalization of
the scalar top and bottom sector at the one-loop level involves a
renormalization of the complex phase $\phiat$. 

Concerning the explicit numerical results we have focused on the
lightest Higgs-boson mass, $\MHe$. 
This is of interest in view of the current exclusion
bounds~\cite{LEPHiggs,Tevbounds} and 
possible high-precision measurements of the properties of a light Higgs
boson at the next generation of colliders~\cite{lhctdrs,LHCrev,ilc}. 
The \order{\alt\als} corrections yield a large downward shift in $\MHe$,
in accordance with the well-known result from the rMSSM.
We find that the impact of the complex phases $\phiat$ and $\phixt$ is 
significantly enhanced by the two-loop contributions, which is a
consequence in particular of diagrams involving internal gluinos. 
We find that varying the complex phases of the scalar top sector 
and of the gluino mass parameter
can induce shifts in $\MHe$ of up to $\sim 5 \gev$ even in cases where the 
one-loop result shows hardly any dependence on the phases. 
The result for $\MHe$ for $\phiat, \phigl \neq 0, \pi$ is found to lie
in the intervals given by $\pm |\At|, \pm |M_3|$. 
The effects of the complex phases of the \order{\alt\als} corrections 
can also be enhanced in the
threshold region where the gluino mass is approximately equal to the sum
of the top-quark mass and the mass of one of the scalar top quarks.

The new results of \order{\alt\als}
will be implemented into the Fortran code
\fh~~\cite{mhiggslong,mhiggsAEC,feynhiggs,mhcMSSMlong}. 
A detailed description, including a comparison of different
renormalization schemes for the scalar top sector and a more elaborate
discussion of Higgs-boson masses and mixings
will be presented in a forthcoming publication~\cite{mhcMSSM2Llong},
as well as
a comparison with the results based on the renormalization-group
improved effective-potential approach~\cite{mhiggsCPXRG2,cpsh}.


\subsection*{Acknowledgements}

We thank T.~Hahn and D.~St\"ockinger for helpful discussions. 
The work of S.H.\ was partially supported by CICYT (grant FPA2006--02315).
Work supported in part by the European Community's Marie-Curie Research
Training Network under contract MRTN-CT-2006-035505
`Tools and Precision Calculations for Physics Discoveries at Colliders'



\begin{thebibliography}{99} 

\bibitem{mhiggsletter} S.~Heinemeyer, W.~Hollik and G.~Weiglein, 
                       {\em Phys. Rev.} {\bf D 58} (1998) 091701, 
                       hep-ph/9803277; 
                       {\em Phys. Lett.} {\bf B 440} (1998) 296, 
                       hep-ph/9807423.

\bibitem{mhiggslong} S.~Heinemeyer, W.~Hollik and G.~Weiglein,
                     {\em Eur. Phys. J.} {\bf C 9} (1999) 343.
                     hep-ph/9812472.

\bibitem{mhiggsFDalbals} S.~Heinemeyer, W.~Hollik, H.~Rzehak and G.~Weiglein,
                         {\em Eur. Phys. J.} {\bf C 39} (2005) 465, 
                         hep-ph/0411114.

\bibitem{bse} S.~Heinemeyer, W.~Hollik and G.~Weiglein,
              {\em Phys. Lett.} {\bf B 455} (1999) 179,
              hep-ph/9903404.
              M.~Carena, H.~Haber, S.~Heinemeyer, W.~Hollik, C.~Wagner,
              and G.~Weiglein,
              {\em Nucl. Phys.} {\bf B 580} (2000) 29,
              hep-ph/0001002.

\bibitem{mhiggsEP} R.~Zhang, 
                   {\em Phys.\ Lett. } {\bf B 447} (1999) 89, 
                   hep-ph/9808299;\\
                   J.~Espinosa and R.~Zhang, 
                   {\em JHEP} {\bf 0003} (2000) 026, 
                   hep-ph/9912236;\\
                   G.~Degrassi, P.~Slavich and F.~Zwirner,
                   {\em Nucl. Phys.} {\bf B 611} (2001) 403,
                   hep-ph/0105096;\\
                   R.~Hempfling and A.~Hoang, 
                   {\em Phys. Lett.} {\bf B 331} (1994) 99, 
                   hep-ph/9401219;\\
                   A.~Brignole, G.~Degrassi, P.~Slavich and F.~Zwirner,
                   {\em Nucl. Phys.} {\bf B 631} (2002) 195,
                   hep-ph/0112177;
                   {\em Nucl. Phys.} {\bf B 643} (2002) 79,
                   hep-ph/0206101;\\
                   J.~Espinosa and R.~Zhang,
                   {\em Nucl. Phys.} {\bf B 586} (2000) 3,
                   hep-ph/0003246;\\
                   J.~Espinosa and I.~Navarro,
                   {\em Nucl.\ Phys.} {\bf B 615} (2001) 82, 
                   hep-ph/0104047;\\
                   G.~Degrassi, A.~Dedes and P.~Slavich,
                   {\em Nucl. Phys.} {\bf B 672} (2003) 144, 
                   hep-ph/0305127.

\bibitem{mhiggsRG1} J.~Casas, J.~Espinosa, M.~Quir\'os and A.~Riotto,
                    {\em Nucl. Phys.} {\bf B 436} (1995) 3,
                    [Erratum-ibid.\ {\bf B 439} (1995) 466],
                    hep-ph/9407389;\\
                    M.~Carena, J.~Espinosa, M.~Quir\'os and C.~Wagner, 
                    {\em Phys. Lett.} {\bf B 355} (1995) 209, 
                    hep-ph/9504316;\\
                    M.~Carena, M.~Quir\'os and C.~Wagner, 
                    {\em Nucl. Phys.} {\bf B 461} (1996) 407, 
                    hep-ph/9508343.

\bibitem{mhiggsAEC} G.~Degrassi, S.~Heinemeyer, W.~Hollik,
                    P.~Slavich and G.~Weiglein, 
                    {\em Eur. Phys. J.} {\bf C 28} (2003) 133,
                    hep-ph/0212020.

\bibitem{PomssmRep} S.~Heinemeyer, W.~Hollik and G.~Weiglein,
                    {\em Phys.\ Rept.} {\bf 425} (2006) 265. 
                    hep-ph/0412214.

\bibitem{mhiggsWN} B.~Allanach, A.~Djouadi, J.~Kneur, W.~Porod and P.~Slavich,
                   {\em JHEP} {\bf 0409} (2004) 044, 
                   hep-ph/0406166.

\bibitem{mhiggsEP5} S.~Martin, 
                    {\em Phys. Rev.} {\bf D 65} (2002) 116003,
                    hep-ph/0111209;
                    {\em Phys. Rev.} {\bf D 66} (2002) 096001,
                    hep-ph/0206136;
                    Phys. Rev. {\bf D 67} (2003) 095012, 
                    hep-ph/0211366;
                    {\em Phys. Rev.} {\bf D 68} 075002 (2003), 
                    hep-ph/0307101; 
                    {\em Phys. Rev.} {\bf D 70} (2004) 016005, 
                    hep-ph/0312092;
                    {\em Phys. Rev.} {\bf D 71} (2005) 016012, 
                    hep-ph/0405022;
                    {\em Phys. Rev.} {\bf D 71} (2005) 116004, 
                    hep-ph/0502168;
                    hep-ph/0701051;\\
                    S.~Martin and D.~Robertson,
                    {\em Comput.\ Phys.\ Commun.} {\bf 174} (2006) 133, 
                    hep-ph/0501132.

\bibitem{mhiggsCPXgen} A.~Pilaftsis,
                       {\em Phys. Rev.} {\bf D 58} (1998) 096010,
                       hep-ph/9803297;
                       {\em Phys. Lett.} {\bf B 435} (1998) 88,
                       hep-ph/9805373.

\bibitem{mhiggsCPXEP} D.~Demir, 
                      {\em Phys. Rev.} {\bf D 60} (1999) 055006,
                      hep-ph/9901389;\\
                      S.~Choi, M.~Drees and J.~Lee,
                      {\em Phys. Lett.} {\bf B 481} (2000) 57,
                      hep-ph/0002287;\\
                      T.~Ibrahim and P.~Nath,
                      {\em Phys. Rev.} {\bf D 63} (2001) 035009, 
                      hep-ph/0008237;
                      {\em Phys. Rev.} {\bf D 66} (2002) 015005,
                      hep-ph/0204092.

\bibitem{mhiggsCPXRG1} A.~Pilaftsis and C.~Wagner, 
                       {\em Nucl. Phys.} {\bf B 553} (1999) 3,
                       hep-ph/9902371.

\bibitem{mhiggsCPXRG2} M.~Carena, J.~Ellis, A.~Pilaftsis and C.~Wagner,
                       {\em Nucl. Phys.} {\bf B 586} (2000) 92,
                       hep-ph/0003180.

\bibitem{mhiggsCPXFD1} S. Heinemeyer,
                       {\em Eur. Phys. J.} {\bf C 22} (2001) 521,
                       hep-ph/0108059.

\bibitem{mhcMSSMlong} M.~Frank, T.~Hahn, S.~Heinemeyer, W.~Hollik, 
                      R.~Rzehak and G.~Weiglein,
                      {\em JHEP} {\bf 02} (2007) 047, 
                      hep-ph/0611326.

\bibitem{dissHR} H.~Rzehak, PhD thesis:
                 ``Two-loop contributions in the supersymmetric Higgs
                 sector'', Technische Universit\"at M\"unchen, 2005; 
                 see: {\tt nbn-resolving.de/} \\
                 with {\tt urn}: {\tt nbn:de:bvb:91-diss20050923-0853568146}~.

\bibitem{feynhiggs} S.~Heinemeyer, W.~Hollik and G.~Weiglein,
                    {\em Comput. Phys. Commun.} {\bf 124} (2000) 76,
                    hep-ph/9812320;
                    hep-ph/0002213;
                    see {\tt www.feynhiggs.de} .

\bibitem{cpsh} J.~Lee, A.~Pilaftsis et al.,
               {\em Comput. Phys. Commun.} {\bf 156} (2004) 283, 
               hep-ph/0307377.

\bibitem{mhcMSSM2Llong} T.~Hahn, S.~Heinemeyer, W.~Hollik, 
                        H.~Rzehak and G.~Weiglein,
                        {\em in preparation}.

\bibitem{Peccei} R.~Peccei and H.~Quinn,
                 {\em Phys.\ Rev.\ Lett.} {\bf 38} (1977) 1440;
                 {\em Phys.\ Rev.} {\bf  D 16} (1977) 1791.

\bibitem{MSSMcomplphasen} S.~Dimopoulos and S.~Thomas,
                          {\em Nucl.\ Phys.} {\bf  B 465} (1996) 23, 
                          hep-ph/9510220.

\bibitem{feynarts} J.~K\"ublbeck, M.~B\"ohm and A.~Denner, 
                   {\em Comp. Phys. Comm.} {\bf 60} (1990) 165;\\
                   T.~Hahn,
                   {\em Comput. Phys. Comm.} {\bf 140} (2001) 418,
                   hep-ph/0012260;\\
                   The program is available via {\tt www.feynarts.de};\\
                   T.~Hahn and C.~Schappacher,
                   {\em Comput. Phys. Comm.} {\bf 143} (2002) 54,
                   hep-ph/0105349.

\bibitem{twocalc} G.~Weiglein, R.~Scharf and M.~B\"ohm,
                  {\em Nucl. Phys.} {\bf B 416} (1994) 606,
                  hep-ph/9310358;\\
G.~Weiglein, R.~Mertig, R.~Scharf and M.~B\"ohm, 
in {\it New Computing Techniques in Physics Research 2},
ed.~D.~Perret-Gallix (World Scientific, Singapore, 1992), p.~617.

\bibitem{oneloop} G.~'t Hooft and M.~Veltman,
                  {\em Nucl. Phys.} {\bf B 153} (1979) 365.

\bibitem{twoloop} A. Davydychev und J.~Tausk, 
                  {\em Nucl. Phys.} {\bf B 397} (1993) 123;\\
                  F. Berends und J.~Tausk,
                  {\em Nucl. Phys.} {\bf B 421} (1994) 456.

\bibitem{looptools} T.~Hahn, M.~Perez-Victoria,
                    {\em Comput.\ Phys.\ Commun.} {\bf 118} (1999) 153, 
                    hep-ph/9807565.

\bibitem{hr} W. Hollik and H. Rzehak, 
             {\em Eur.\ Phys.\ J.} {\bf C 32} (2003) 127, 
             hep-ph/0305328.

\bibitem{SUSYphases} M.~Dugan, B.~Grinstein and L.~Hall,
                     {\em Nucl.\ Phys.} {\bf B 255} (1985) 413.

\bibitem{pdg} W.~Yao et al.\  [Particle Data Group Collaboration],
              {\em J.\ Phys.} {\bf G 33} (2006) 1.

\bibitem{plehnix} V.~Barger, T.~Falk, T.~Han, J.~Jiang, T.~Li and T.~Plehn,
                  {\em Phys. Rev.} {\bf D 64} (2001) 056007, 
                  hep-ph/0101106.

\bibitem{EDMDoink} W.~Hollik, J.~Illana, S.~Rigolin and D.~St\"ockinger,
                   {\em Phys. Lett.} {\bf B 416} (1998) 345, 
                   hep-ph/9707437;
                   {\em Phys. Lett.} {\bf B 425} (1998) 322, 
                   hep-ph/9711322.

\bibitem{EDMrev2} D.~Demir, O.~Lebedev, K.~Olive, M.~Pospelov and A.~Ritz,
                  {\em Nucl. Phys.} {\bf B 680} (2004) 339, 
                  hep-ph/0311314.

\bibitem{EDMPilaftsis} D.~Chang, W.~Keung and A.~Pilaftsis,
                       {\em Phys. Rev. Lett.} {\bf 82} (1999) 900
                       [Erratum-ibid.\  {\bf 83} (1999) 3972], 
                       hep-ph/9811202;\\
                       A.~Pilaftsis,
                       {\em Phys. Lett.} {\bf B 471} (1999) 174, 
                       hep-ph/9909485.

\bibitem{EDMRitz} O.~Lebedev, K.~Olive, M.~Pospelov and A.~Ritz,
                  {\em Phys. Rev.} {\bf D 70} (2004) 016003, 
                  hep-ph/0402023.

\bibitem{LEPHiggs} [LEP Higgs working group],
                   {\em Phys. Lett.} {\bf B 565} (2003) 61,
                   hep-ex/0306033;
                   {\em Eur.\ Phys.\ J.} {\bf C 47} (2006) 547,
                   hep-ex/0602042.

\bibitem{Tevbounds} V.~Abazov et al.  [D0 Collaboration],
                    {\em Phys.\ Rev.\ Lett.} {\bf 97} (2006) 121802, 
                    hep-ex/0605009;
                    D0~Note 5331-CONF;\\
                    A.~Abulencia et al.  [CDF Collaboration],
                    {\em Phys.\ Rev.\ Lett.} {\bf 96} (2006) 011802,
                    hep-ex/0508051;
                    CDF note 8676;\\
                    {}[CDF Collaboration],
                    {\em Phys.\ Rev.\ Lett.} {\bf 96} (2006) 042003,
                    hep-ex/0510065.

\bibitem{lhctdrs} ATLAS Collaboration,
        {\em Detector and Physics Performance Technical Design Report},
        CERN/LHCC/99-15 (1999), see:\\
        {\tt atlasinfo.cern.ch/Atlas/GROUPS/PHYSICS/TDR/access.html} ;\\
        CMS Collaboration,
        {\em Physics Technical Design Report, Volume 2. CERN/LHCC
          2006-021}, 
        see: {\tt cmsdoc.cern.ch/cms/cpt/tdr/} .

\bibitem{LHCrev} V.~B\"uscher and K.~Jakobs,
                 {\em Int.\ J.\ Mod.\ Phys.} {\bf A 20} (2005) 2523,
                 hep-ph/0504099;\\
                 M.~Schumacher,
                 {\em Czech. J. Phys.} {\bf 54} (2004) A103;
                 hep-ph/0410112.

\bibitem{ilc} J.~Aguilar-Saavedra et al.,
              TESLA TDR Part~3: 
              ``Physics at an $e^+e^-$ Linear Collider'', 
              hep-ph/0106315,
              see: {\tt tesla.desy.de/tdr/};\\
              T.~Abe et al.  
              [American Linear Collider Working Group Collaboration],
              hep-ex/0106056;\\
              K.~Abe et al. 
              [ACFA Linear Collider Working Group Collaboration],
              hep-ph/0109166;\\
              K.~Ackermann et al.,
              DESY-PROC-2004-01;\\
              S.~Heinemeyer et al.,
              hep-ph/0511332.

\end{thebibliography}
\end{document}